\documentclass[journal,10pt,twocolumn,twoside]{IEEEtran}
\usepackage{ifpdf}
\usepackage{cite}
\usepackage{amsmath,amssymb,amsfonts,bm}
\usepackage{algorithmic}
\usepackage{graphicx, adjustbox}
\usepackage{array}
\usepackage{enumitem}
\usepackage{breqn}
\usepackage{bm}
\usepackage{dsfont}
\usepackage{upgreek}
\usepackage{textcomp}
\usepackage{multirow}
\usepackage{algorithm}
\usepackage{caption}
\usepackage{pgfplots}
\usepackage{tkz-euclide}
\usepackage{subcaption}
\usepackage{tikz}
\usepackage{hyperref}
\usepackage{balance}
\usepackage{tipa}

\usepackage{upgreek}
\usepackage{soul}
\usepackage{breqn}
\usepackage{makecell}
\usepackage{tabularray}
\usepackage[english]{babel}

\addto\captionsenglish{}
\allowdisplaybreaks

\usepackage{amsthm}

\usepackage{fancyhdr}

\usepackage{graphicx}
\usepackage[table]{xcolor} 

\makeatletter
\newcommand{\vast}{\bBigg@{4}}

\newcommand{\Vast}{\bBigg@{5}}
\makeatother

\pgfplotsset{compat=1.18}
\definecolor{Gray}{gray}{0.9}
\definecolor{darkyellow}{RGB}{249, 231, 159} 
\definecolor{salmon}{RGB}{255, 228, 213}
\definecolor{darksalmon}{RGB}{219, 195, 181}
\definecolor{darkgreen}{RGB}{180, 230, 198}
\definecolor{lightgreen}{RGB}{209, 233, 201}

\definecolor{lightred}{RGB}{255,211,222}
\definecolor{darkred}{RGB}{156, 66, 45}
 \definecolor{LightBlue}{rgb}{0.8,0.89,1} 
\definecolor{DarkBlue}{rgb}{0.57,0.71,0.82} 
\definecolor{MagLight}{rgb}{1, 0.89, 0.8}
\usepackage[font={small},labelfont=bf]{caption}
\usepackage{enumitem}
  \makeatletter
\newcommand\notsotiny{\@setfontsize\notsotiny\@vipt\@viipt}
\makeatother
\usepackage[most]{tcolorbox}
\begin{document}

\title{Empowering AI-Native 6G Wireless Networks with Quantum Federated Learning}

\author{{
Shaba Shaon, Md Raihan Uddin, Dinh C. Nguyen, Seyyedali Hosseinalipour,~\textit{Senior Member, IEEE}, \\ Dusit Niyato,~\textit{Fellow, IEEE},  Octavia A. Dobre,~\textit{Fellow, IEEE}

}

\thanks{Shaba Shaon, Md Raihan Uddin, Dinh C. Nguyen are with the Department of Electrical and Computer Engineering, The University of Alabama in Huntsville, USA. Emails: ss0670@uah.edu, mu0016@uah.edu, dinh.nguyen@uah.edu.}
\thanks{Seyyedali Hosseinalipour is with the Department of Electrical Engineering, University at Buffalo--SUNY, NY, USA. (e-mail: alipour@buffalo.edu).}
\thanks{Dusit Niyato is with the College of Computer Science and Engineering, Nanyang Technological University, Singapore (e-mail: dniyato@ntu.edu.sg).}
\thanks{Octavia Dobre is with the Faculty of Engineering and Applied Science, Memorial University, Canada (e-mail: odobre@mun.ca).}

}

{}

\maketitle

\begin{abstract}
AI-native 6G networks are envisioned to tightly embed artificial intelligence (AI) into the wireless ecosystem, enabling real-time, personalized, and privacy-preserving intelligence at the edge. A foundational pillar of this vision is federated learning (FL), which allows distributed model training across devices without sharing raw data. However, implementing classical FL methods faces several bottlenecks in heterogeneous dynamic wireless networks, including limited device compute capacity, unreliable connectivity, intermittent communications, and vulnerability to model security and data privacy breaches.
This article investigates the integration of quantum federated learning (QFL) into AI-native 6G networks, forming a transformative paradigm capable of overcoming these challenges. By leveraging quantum techniques across computing, communication, and cryptography within FL workflows, QFL offers new capabilities along three key dimensions: (i) edge intelligence, (ii) network optimization, and (iii) security and privacy, which are studied in this work.  We further present a case study demonstrating that a QFL framework employing the \textcolor{black}{quantum approximate optimization algorithm} outperforms classical methods in model convergence. We conclude the paper by identifying practical challenges facing QFL deployment, such as quantum state fragility, incompatibility with classical protocols, and hardware constraints, and then outline key research directions toward its scalable real-world adoption.

\end{abstract}

\begin{IEEEkeywords}
Quantum federated learning, 6G networks, network optimization, quantum computing and communications.
\end{IEEEkeywords}

\section{Introduction}

AI-native 6G wireless networks are envisioned to transform communication systems by integrating artificial intelligence (AI) into the network infrastructure, enabling real-time execution of tasks such as adaptive resource allocation, autonomous control, and personalized service delivery at the network edge. 
A key enabler of this vision is \textit{federated learning (FL)}, a distributed machine learning (ML) framework that relies solely on the exchange of devices' local model parameters during the model training phase, thereby reducing communication overhead and mitigating the privacy risks associated with raw data transmission over the network~\cite{nguyen20216g}. Nevertheless, the adoption of conventional FL frameworks faces various hurdles: limited compute capacity of edge devices, dynamic network conditions leading to intermittent communications, and persistent privacy and security risks that arise even when only model parameters are exchanged. These challenges can hinder the integration of \textit{distributed intelligence} within AI-native 6G networks.

\begin{figure}[ht!]
    \centering
    \includegraphics[width=0.99\linewidth]{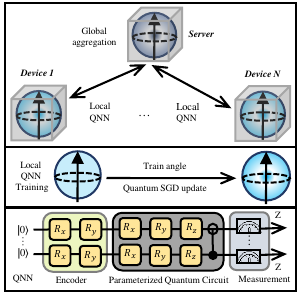}
    \caption{Schematic of a QFL framework. Each device's local QNN consists of a state encoder and a \textcolor{black}{parameterized quantum circuit}, which has trainable  parameters. The measurement outcomes are used to update the QNN parameters via a classical optimizer (e.g., SGD). The local models are then sent to the server for aggregation using secure protocols, \textcolor{black}{such as post-quantum
cryptography (PQC).}}
    \label{Fig:Overview}
\end{figure}

A promising paradigm that can address these challenges is \textit{quantum federated learning (QFL)}, which leverages recent technological advancements in quantum computing, quantum communication, and quantum cryptography.  QFL harnesses these quantum technologies to accelerate model training, enhance the efficiency of model parameter transmissions across the network, and strengthen security and privacy protections (e.g., against  model inversion/reconstruction attacks), making its integration into AI-native 6G networks highly enticing~\cite{quy2024federated}.

Schematic of a realization of QFL is illustrated in Fig.~\ref{Fig:Overview}, where distributed devices collaborate with a server to train a global model. Specifically, each device first encodes its local data into quantum states using a quantum state encoder, then processes the input through a local quantum neural network (QNN) constructed from \textcolor{black}{a \textit{parameterized quantum circuit} (the core of the QNN model shown in the figure)}. Measurement outcomes from the circuit are then used to compute gradients, and the model is updated locally (e.g., using classical optimization techniques such as stochastic gradient descent (SGD)). The updated model parameters are then transmitted to the server using \textcolor{black}{secure protocols, such as \textit{post-quantum cryptography (PQC)}}, where secure aggregation is performed at the server to construct the global model. \textcolor{black}{An added feature of this communication process, distinct from classical FL, is the nature of the exchanged information. Instead of transmitting the potentially large weight tensors of a deep neural network, QFL clients typically send a compact and smaller vector of classical parameters representing the trainable angles of the parameterized quantum circuit.} This global model is subsequently broadcasted back to all devices, enabling the next round of collaborative training.  
It is important to note that this is just one example of QFL realization; there is no unified definition of QFL or its operations due to its emerging nature. For instance, some QFL frameworks may retain classical learning and transmission operations while leveraging quantum-inspired network optimization techniques for device orchestration and scheduling.

\begin{tcolorbox}[title={Paper Overview and Vision}, colback=gray!5, colframe=darkred, fonttitle=\bfseries]
This paper presents a pioneering exploration of the transformative potentials of QFL in advancing three core pillars of AI-native 6G: (i) edge intelligence, (ii) network optimization, and (iii) security and privacy. We also investigate the benefits of QFL by presenting a representative case study. Furthermore, we identify critical challenges associated with the deployment of QFL in  6G environments and outline promising research directions aimed at bridging the gap between theoretical advances and the real-world implementation of QFL.

$\star$ Given that many quantum-empowered algorithms remain largely at the theoretical stage and their specific applications to QFL operations have not been thoroughly studied, we intentionally adopt a speculative tone in several parts of this paper to reflect the theoretical nature of the claims.
\end{tcolorbox}

\begin{figure*}[ht!]
\centerline{\includegraphics[width=0.99\linewidth]{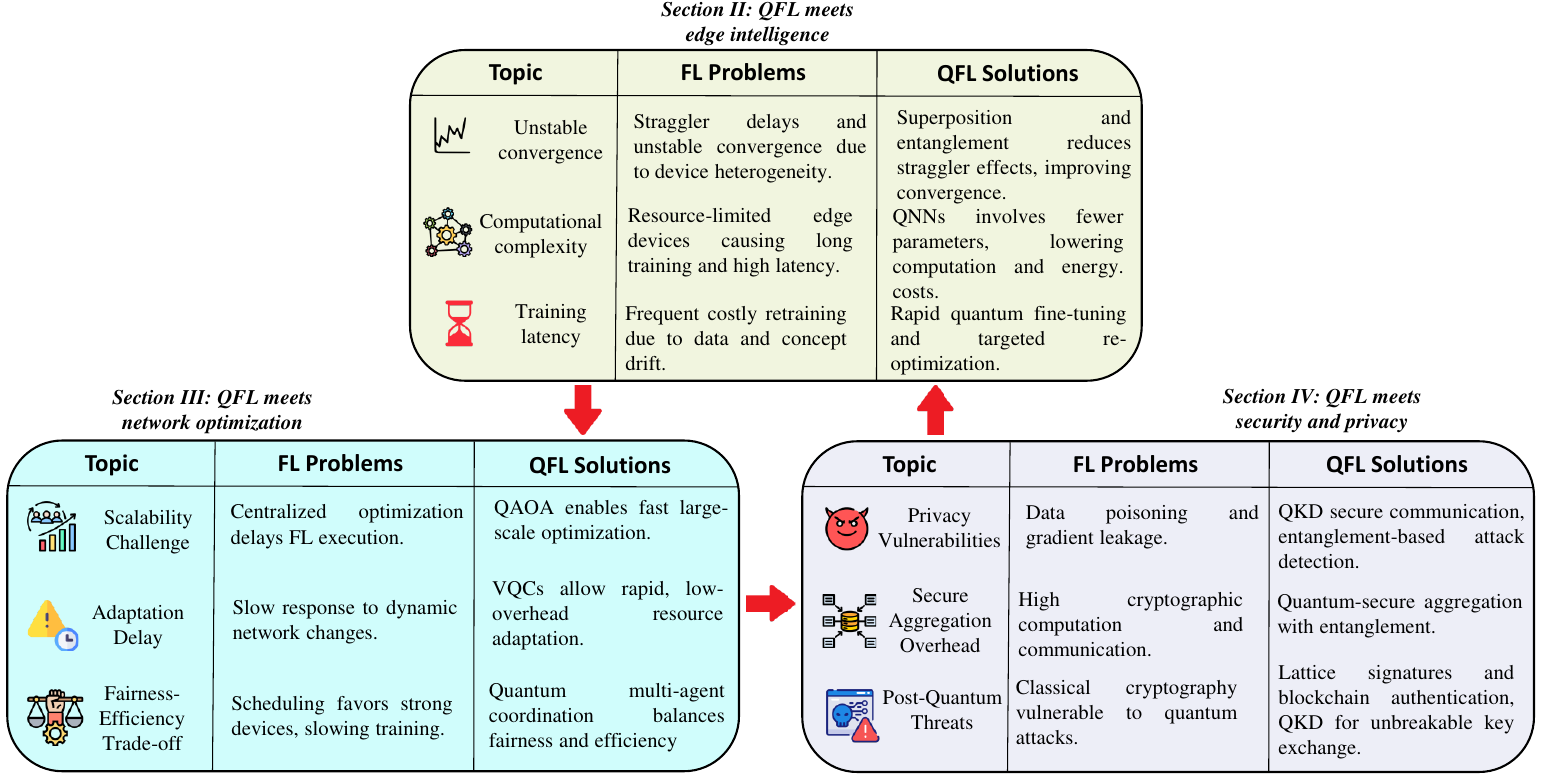}}
\caption{Summary of key challenges in FL across edge intelligence, network optimization, and security/privacy domains, alongside QFL solutions. The top box outlines FL challenges related to convergence, computational complexity, and training latency, with corresponding QFL enhancements. The bottom-left box addresses network scalability, adaptation delays, and fairness-efficiency trade-offs, highlighting quantum algorithms such as QAOA and VQCs as effective solutions. The bottom-right box focuses on privacy vulnerabilities, secure aggregation overhead, and emerging post-quantum threats, illustrating QFL’s use of QKD, quantum-secure aggregation, and post-quantum cryptographic methods to safeguard distributed intelligence in AI-native 6G networks.}
\label{fig1}
\end{figure*}

\section{QFL Meets Edge Intelligence: Enhancing Model Training Speed}

In this section, we outline three key challenges associated with implementing conventional FL at the wireless edge and detail how the integration of QFL into AI-native 6G networks can overcome them.

\subsection{Convergence Bottlenecks of FL  in Heterogeneous Wireless Networks}

Conventional FL faces notable convergence bottlenecks in wireless networks due to the inherent heterogeneity of edge devices. These systems often comprise a mix of devices with \textit{varying computational power, memory availability, energy budgets, and communication capabilities}. In conventional FL with synchronous global aggregations, this heterogeneity manifests in the form of \textit{straggler effects}, where slower/resource-limited devices delay each global aggregation round, prolonging training and reducing the overall system efficiency. Asynchronous FL variants aim to address this by allowing devices to upload their updated models independently/sequentially; however, this can introduce stale or inconsistent updates into the global model  and may destabilize its convergence, particularly when data is highly non-identically distributed (non-IID) across devices. These issues are further exacerbated in \textit{dynamic} wireless environments, where devices are mobile, intermittently connected, and progress through training at differing rates. Under such conditions, collecting consistent model updates from devices becomes increasingly difficult, and global model convergence is often  compromised.

$\star$ \textbf{ \textcolor{black}{The QFL Solution:}} 
QFL offers a promising path to overcoming these bottlenecks by making the model training and its results more equitable across devices (e.g., even weaker devices can contribute  meaningful model updates to the global model, preventing the global model from being biased towards powerful devices). In particular, QFL can integrate quantum-enhanced optimization methods such as the \textit{quantum approximate optimization algorithm (QAOA)} to reduce the number of iterations required to obtain local model updates. These methods leverage quantum principles such as \textit{superposition} and \textit{entanglement} to explore multiple parameter configurations in parallel, potentially improving learning efficiency under constrained computational resources. Also, QFL can employ hybrid optimization strategies such as variational quantum algorithms \cite{liu2022layer}, where QNNs are trained using classical gradient descent applied to parameterized quantum circuits. In these settings, quantum circuits process and encode data in a high-dimensional Hilbert space, while gradients are computed via techniques such as the \textit{parameter-shift rule}, an approach referred to as quantum gradient descent. This reduces the effective parameter space dimensionality and achieves faster model convergence.

\subsection{Complexity and Computational Strain of FL at the Edge}
    


For resource-constrained edge devices, the deployment and training of conventional ML models (especially deep neural networks) in FL can impose significant computational demands. Consequently, this leads to several concerns: extended local training time, accelerated battery drain, reduced device participation, and increased overall system-level latency. While techniques such as model compression, split learning, and hierarchical FL have been proposed to mitigate these issues, they introduce additional coordination overhead and communication/computation complexity, which may not scale well in large-scale wireless deployments.

$\star$ \textbf{The QFL Solution:} \textcolor{black}{QFL addresses these overheads through the unique properties of the underlying quantum technology. From the communication standpoint, the overhead could be decreased since clients can transmit only a compact vector of classical parameters that define the parameterized quantum circuit instead of the entire parameter set of a deep neural network. From the computation standpoint,} QNNs exploit quantum principles such as superposition and entanglement to represent and process information more \textit{compactly} in high-dimensional Hilbert spaces. This enables hybrid quantum-classical models to often achieve high accuracy with significantly fewer trainable parameters and shallower circuit depths compared to their classical counterparts~\cite{quy2024federated}. \textcolor{black}{As a result, while executing a quantum circuit and its classical optimizer still demands local resources, the model's compactness and potential for faster convergence can ultimately reduce the overall training time and energy consumption on edge devices, mitigating the primary strains of classical FL.}

\subsection{(Re)Training Latency of FL  under Data and Concept Drift}
 In wireless networks, edge devices continuously collect data that evolves over time due to shifting user behavior, environmental dynamics, and variability in sensor performance. These changes give rise to \textit{data drift} (i.e., changes in input/feature distributions over time) and \textit{concept drift} (i.e., changes in the relationship between inputs/features and outputs/labels over time).  If not properly addressed through continual model adaptation, such drifts can degrade the stability and accuracy of the global model.  FL approaches typically rely on traditional continual learning methods that often require frequent model retraining or complex memory management, which are computationally costly for resource-constrained devices. 

$\star$ \textbf{\textcolor{black}{The QFL Solution:}} QFL offers a promising path to address data and concept drift by enabling more rapid model adaptation at the devices. In particular, QFL can incorporate quantum-enhanced optimization methods, such as the QAOA or quantum annealing \cite{choi2019tutorial, ye2023quantum}, to accelerate local model fine-tuning when new data distributions emerge. Rather than retraining the entire model, these algorithms can be used to re-optimize targeted subcomponents: for example, QAOA can be applied to adjust discrete layer-wise hyperparameters (e.g., dropout rates or activation quantization levels) by framing them as combinatorial optimization problems. Also, in classification tasks, the model's task head (i.e., the final layer responsible for mapping learned representations to output labels) can be re-calibrated using quantum annealing to efficiently update model weights in response to changes in class/label distributions.



\section{QFL Meets Network Optimization: Improving Resource Allocation and Device Scheduling}
In this section, we explore three key network optimization/orchestration challenges facing the implementation of FL over large-scale dynamic wireless networks, and  demonstrates how QFL's integration into AI-native 6G networks offers effective solutions.



\subsection{Scalability Challenges of FL}

FL implementation is intrinsically coupled with network optimization. In particular, the smooth execution of FL requires efficient optimization of (i) device sampling/scheduling, (ii) local training configurations (e.g., SGD mini-batch size, number of SGD iterations, CPU clock frequency/speed), and (iii) model/gradient transmission (e.g., bandwidth allocation and transmit power control). These can be collectively thought of as optimization variables that must be tuned for efficient FL execution under various objective functions, such as FL training latency or energy minimization. 

To solve these often non-convex optimization problems, FL typically relies on classical convex/non-convex optimization techniques. These techniques often involve iterative procedures (e.g., gradient descent) to find solutions, which can delay the execution of FL operations. Moreover, these optimizations are typically formulated and solved centrally (e.g., at a cloud/edge server), where the solution space scales with the number of devices, causing the FL execution delay to grow with the network size.
Although distributed optimization techniques can be implemented among devices, these methods are still iterative and further rely on information exchange among devices that can incur a high communication overhead.

$\star$ \textbf{\textcolor{black}{The QFL Solution:}} 
QFL can address these bottlenecks by leveraging the QAOA to rapidly solve large-scale optimization problems. In particular, QAOA iteratively refines variational parameters through hybrid quantum-classical loops by encoding optimization variables (e.g., device scheduling and bandwidth allocation) into a cost Hamiltonian, enabling near-instantaneous solutions. For instance, in the broader network optimization domain, the \textit{quantum adaptive task and optimization scheduler (QATOS)} framework is shown to be capable of fine-tuning quantum circuit parameters and achieving efficient load balancing across heterogeneous wireless nodes by dynamically adjusting resource allocations in real-time using QAOA \cite{choi2019tutorial}.

\subsection{Adaptation Delay of FL in Dynamic Networks}
Besides the network scale discussed above, real-time device mobility and dynamic channel conditions further complicate the network optimization in FL. In particular, conventional optimization techniques used in FL often struggle to respond quickly to real-time mobility and changing channel conditions. On one hand, fixed resource allocation solutions (e.g., optimized once before the execution of FL and maintained throughout the  FL training period) are unable to adapt to variations in wireless channel quality, resulting in degraded model accuracy and increased energy consumption and latency. On the other hand, (re)solving the network optimizations introduces excessive delays prior to the execution of each FL round, which can accumulate into significant latency over the entire FL training window. 
While network dynamics can be predicted using learning-based methods (e.g., recurrent neural networks), these predictions might fail to match real-time network conditions due to the unpredictable mobility of wireless nodes, leading to suboptimal solutions.

$\star$ \textbf{\textcolor{black}{The QFL Solution:}} 
QFL can leverage \textit{variational quantum circuits (VQCs)} to obtain fast low-overhead solutions that dynamically respond to mobility and channel fluctuations in wireless networks. In particular, the circuit parameters of each VQC can be iteratively optimized through hybrid quantum-classical loops, taking into account factors such as device capabilities and channel conditions. This enables faster adaptation to changing edge conditions: the optimized VQC parameters can dynamically adjust resource allocations (e.g., bandwidth and device scheduling) with potentially reduced overhead compared to  classical optimization schemes. For example, a \textit{quantum deep reinforcement learning (QDRL)} framework, demonstrated in simulations where VQCs allocate wireless resources in 6G networks, is shown to achieve equivalent performance to classical deep reinforcement learning with less parameter complexity and latency \cite{quy2024federated}. By further leveraging the inherent parallelism (i.e., the ability to represent and explore multiple resource allocation configurations simultaneously through quantum superposition)  of VQCs, QFL systems can enable low-latency adaptations to dynamic wireless networks.


\subsection{Fairness-Efficiency Trade-offs of FL: Addressing the ``Straggler" Effect} 

Resource allocation among devices, \textit{in particular device scheduling}, in conventional FL involves a trade-off between maximizing the global model convergence speed for learning/prediction efficiency and ensuring meaningful participation from less capable devices (in terms if computation/communication capabilities) to maintain the model's fairness (i.e., the global model should not become biased towards the dataset of capable devices due to lack of participation of lower-capable devices). On one hand, simple resource allocation rules favoring devices withe better capabilities (i.e., more frequent scheduling of devices with better channel conditions and faster CPU clock cycles) could speed training but marginalize less capable devices, producing models with poor generalization. On the other hand, enforcing strict fairness by equally weighting the participation of all devices can greatly slow down the execution of FL because of the inclusion of stragglers (i.e., less capable devices), compromising real-time wireless applications (e.g., real-time traffic prediction for autonomous vehicles). Adaptive device scheduling methods improve fairness by considering device resources and accuracy bias jointly; however, they often increase training time without careful tuning. For example, Meta's q-Fair FL adds a tunable parameter to balance accuracy and loss \cite{zhou2023resource}, yet optimal values of this parameter are application-specific and non-trivial to obtain.


$\star$ \textbf{\textcolor{black}{The QFL Solution:}} 
QFL offers a promising solution by transforming device scheduling into a dynamic multi-objective optimization problem, leveraging VQCs. Central to this approach is the concept of modeling each edge device as a quantum agent, enabling adaptive multi-agent coordination through variational quantum policies. In particular, shared entanglement among these quantum agents facilitates joint scheduling decisions while reducing the heavy communication overhead typically associated with distributed optimization. For instance, in an entangled quantum multi-agent reinforcement learning setup \cite{ryu2025multi}, agents coupled via quantum channels have demonstrated significant reductions in the number of parameters that must managed for coordination, accelerating consensus. Furthermore, hybrid quantum-classical actor-critic systems \cite{adeniyi2024reinforcement} harness quantum superposition and the inherent parallelism of VQCs to rapidly explore a vast solution space of potential scheduling strategies. This enables nuanced straggler management techniques. For example, the VQC can learn to assign less computationally intensive tasks to less capable devices (e.g., reducing the number of SGD iterations), or strategically schedule them during periods of lower network load, ensuring their valuable model contributions are incorporated into the global model without compromising overall training speed.

\section{QFL Meets Security and Privacy: Enhancing the Robustness of Information Transmission}
In this section, we examine three critical security and privacy challenges in FL  and discuss how integrating QFL into AI-native 6G networks can address them.


\begin{figure}[ht!]
\centerline{\includegraphics[width=0.99\linewidth]{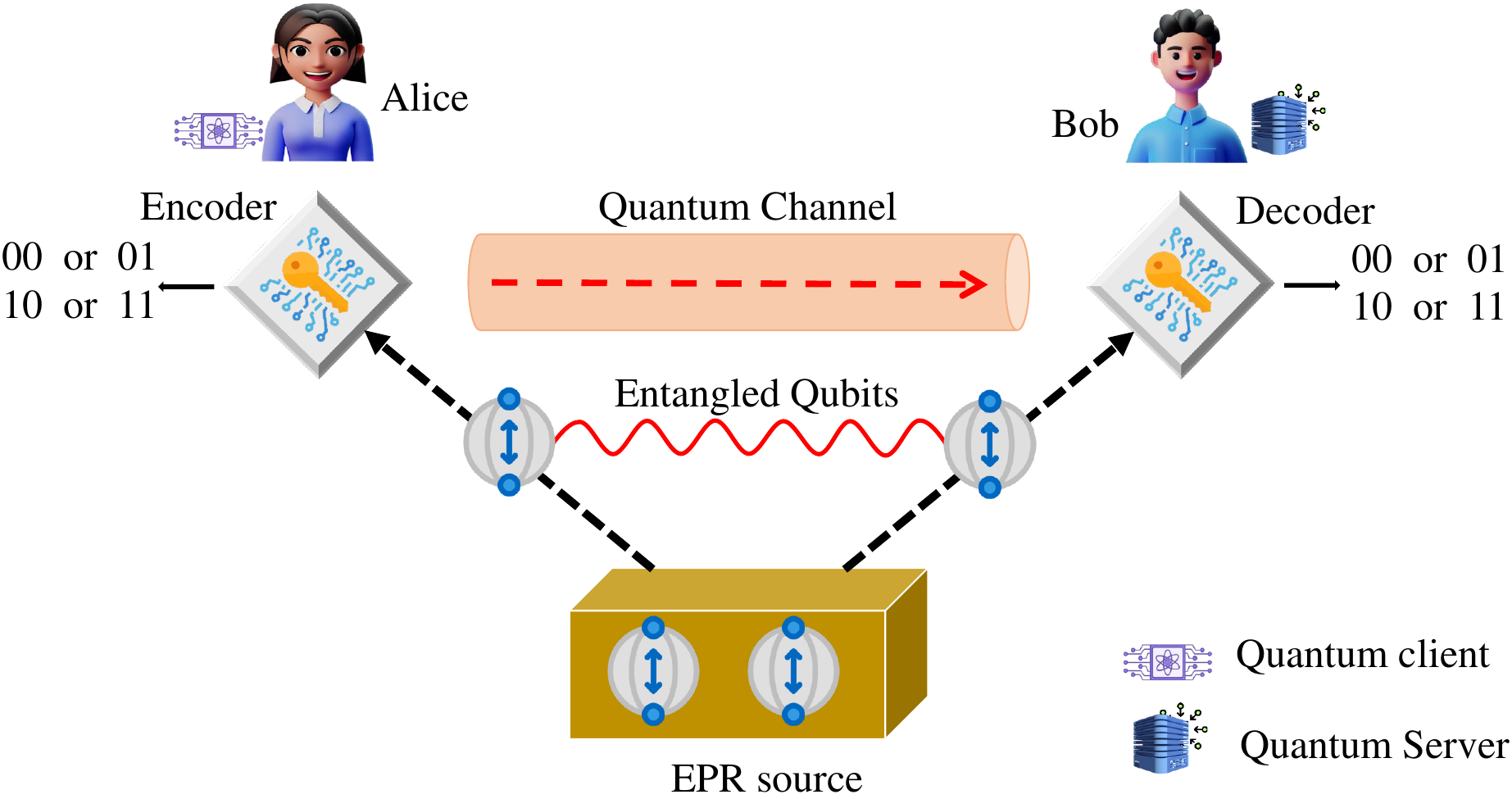}}
\caption{Secure model sharing in QFL between the quantum device (Alice) and the quantum server (Bob) via entanglement. An \textit{Einstein-Podolsky-Rosen (EPR)} source generates pairs of entangled qubits that are distributed to both Alice and Bob. These entangled qubits enable correlated quantum communication through the quantum channel, allowing Alice to securely encode model information that only Bob can decode.}
\label{fig:dense}
\end{figure}

\subsection{Privacy Vulnerabilities of FL}
Conventional FL is vulnerable to various attacks. For instance, in \textit{data poisoning attacks}, adversaries subtly manipulate the local datasets of devices to degrade the performance of the resulting global model. In \textit{backdoor attacks}, an attacker implants specific patterns (known as \textit{triggers}) into training data, causing the model to associate these triggers with adversary-specified outputs. Moreover, \textit{gradient leakage attacks} (e.g., model inversion and reconstruction attacks) can exploit the transmitted devices' gradients during model aggregations to reconstruct sensitive device data. Collectively, these vulnerabilities erode device trust and reduce their willingness to participate in collaborative training.

$\star$ \textbf{The QFL Solution:} 
To protect against classical threats such as gradient leakage, \textit{quantum key distribution (QKD)}, integrated in QFL, provides provably secure key exchange to safeguard communication channels. Any eavesdropping attempt inevitably disturbs the quantum states, allowing for immediate intrusion detection~\cite{wei2023quantum}. \textcolor{black}{Also, to counter data poisoning and backdoor attacks, quantum properties can be leveraged: for example,} using quantum entanglement to correlate device model updates could provide a mechanism to detect malicious or inconsistent model contributions, \textcolor{black}{as an anomalous model update would break the expected quantum correlations.} While the practical realization of such entanglement-based defenses remains an active area of research, it highlights a promising direction for building more robust distributed learning frameworks.
\textcolor{black}{It is worth mentioning that, while introducing these novel defense mechanisms, the integration of quantum elements in QFL also expands the attack surface, creating novel vulnerabilities that must be considered. For example, adversaries can develop \textit{adversarial quantum states} (i.e., subtly perturbed inputs designed to fool a QNN) or launch attacks targeting the integrity of the quantum computation itself, such as manipulating the classical parameters of a hybrid model to compromise its quantum components. The sensitivity of quantum hardware to its environment can also be exploited through fault-injection or side-channel attacks. These emerging threats necessitate the development of quantum-aware adversarial models, robust training algorithms, and hardware-level protections to ensure end-to-end security and privacy in QFL.}

\subsection{Secure Aggregation in FL: Communication and Computation Overhead}

In the above-mentioned attack scenarios, the attacker either manipulates the data or intercepts the devices' transmitted models/gradients by eavesdropping on the communication channels. Nevertheless, another class of threats also falls under the umbrella of gradient leakage attacks, where an adversary gains unauthorized access to the server (e.g.,  by breaching its firewall) and obtains both the transmitted device models/gradients and the aggregated global model. The attacker can then exploit the collected gradients to reconstruct sensitive training data or infer membership information about individual data samples, highlighting the need for \textit{secure model aggregation}.
However, achieving secure aggregation in FL often incurs significant computational and communication overhead, primarily because aggregation must be performed without revealing individual device contributions, necessitating complex cryptographic techniques. Methods such as \textit{functional encryption} and \textit{Google's SecAgg} \cite{nazemi2024boosting}, based on secure multi-party computation principles such as secret sharing and additive homomorphism, rely on intricate mathematical operations that are resource-intensive to execute. For instance, each device must perform demanding cryptographic computations, including generating secret shares, encrypting gradients, and creating blinding masks before transmitting their updates. Meanwhile, the server must engage in cryptographic operations to securely construct the aggregated model. Beyond the computational burden, these protocols often impose additional rounds of information exchange at devices, which can increase synchronization delays and consume extra bandwidth.

$\star$ \textbf{\textcolor{black}{The QFL Solution:}} 
QFL can leverage quantum-secure aggregation (QSA) schemes, where devices encode their models/gradients into quantum states for transmission. These schemes, depicted in Fig.~\ref{fig:dense}, are designed such that upon specific quantum operations and measurement, only the aggregated sum is collectively revealed at the server to perform model aggregation, while devices' contributions remain computationally or information-theoretically private. For instance, by leveraging quantum entanglement, quantum-safe secret sharing protocols can correlate device updates, enabling a form of collective masking where quantum correlations ensure that only the aggregate sum is accessible at the server.  Furthermore, recent advances propose utilizing homomorphic pseudorandom functions on quantum states to enable highly efficient in-network masking and unmasking mechanisms, which are theorized to lower the communication overhead compared to other secure aggregation methods \cite{qu2025daqfl}.

\subsection{Secure Aggregation in FL: Post-Quantum Threats}

Beyond the above-discussed communication and computation overheads of FL-specific secure aggregation protocols, FL may fundamentally depend on classical public-key cryptographic schemes such as \textit{Rivest–Shamir–Adleman (RSA)} and \textit{elliptic-curve cryptography (ECC)} to secure the transmission of devices' models/gradients, especially when FL is executed over conventional wireless networks that may not support/allow FL-specific encryption schemes. In particular, many FL scenarios will rely on RSA and ECC since these techniques are established in  standard secure communication protocols, such as \textit{transport layer security (TLS)}. 
However, advances in quantum computing introduce a critical threat: the danger of rendering classical encryption methods obsolete. Specifically, public-key crypto-systems such as RSA and ECC are vulnerable to \textit{Shor's algorithm}, which enables a powerful quantum computer to factor large integers and solve discrete logarithm problems in polynomial time. Consequently, national cybersecurity authorities are advising organizations to transition to quantum-resilient cryptography, anticipating `Q-Day' (i.e., the hypothetical point when quantum computers become capable of compromising current encryption) could arrive as early as 2035.

$\star$ \textbf{The QFL Solution:} 
To address the above threat, QFL can leverage a two-layer security approach. First, it can use the national institute of standards and technology (NIST)-standardized \textit{PQC algorithms} (e.g., Dilithium, FALCON) to ``future-proof" the classical communication and authentication layers. \textcolor{black}{However, this transition presents its own challenges: for example, PQC algorithms often come with larger key sizes and higher computational overhead compared to their classical counterparts, which can strain the resources of edge devices and potentially increase the overall training latency. This trade-off between post-quantum security and performance is a critical risk that must be studied and managed in 6G deployments.} Second, for enhanced security, QFL can combine these PQC schemes with quantum communication techniques: for example, protocols such as \textit{post-quantum secure blockchain-based federated learning (PQS-BFL)} can use quantum-resistant signatures to authenticate model updates~\cite{qu2025daqfl}. Concurrently, QKD can be used for session key exchange, providing secrecy based on quantum principles. \textcolor{black}{Nevertheless, it is important to recognize that QKD also may face practical limitations in mobile 6G networks, including distance constraints, high infrastructure costs, and vulnerability to denial-of-service (DoS) attacks on the quantum channel. Therefore, QFL does not eliminate \textit{all} post-quantum risks but rather provides a set of complementary solutions that can be carefully engineered into a holistic security architecture.}

\section{Case Study}
\textcolor{black}{Given that dedicated quantum resources will likely remain scarce and costly in the near-term, it is crucial to identify 6G applications where QFL can provide the most significant benefits. The most promising candidates are computationally intensive optimization problems, particularly those that are NP-hard and hinder the performance of the network at scale. Key examples include: (i) dynamic radio resource management (e.g., joint channel allocation and beamforming in massive MIMO systems), (ii) end-to-end network slice orchestration, and (iii) complex scheduling for ultra-reliable low-latency communications (URLLC) services. In these problems, the difficulty of finding optimal solutions with classical algorithms creates performance bottlenecks. By leveraging quantum algorithms for these specific problems, QFL can unlock substantial gains in network efficiency that justify the use of quantum hardware. Below, we tackle such problems to demonstrate the potential of QFL.}

\begin{figure*}[ht!]
    \centering
    \begin{subfigure}[b]{0.48\linewidth}
        \centering
        \includegraphics[width=\linewidth]{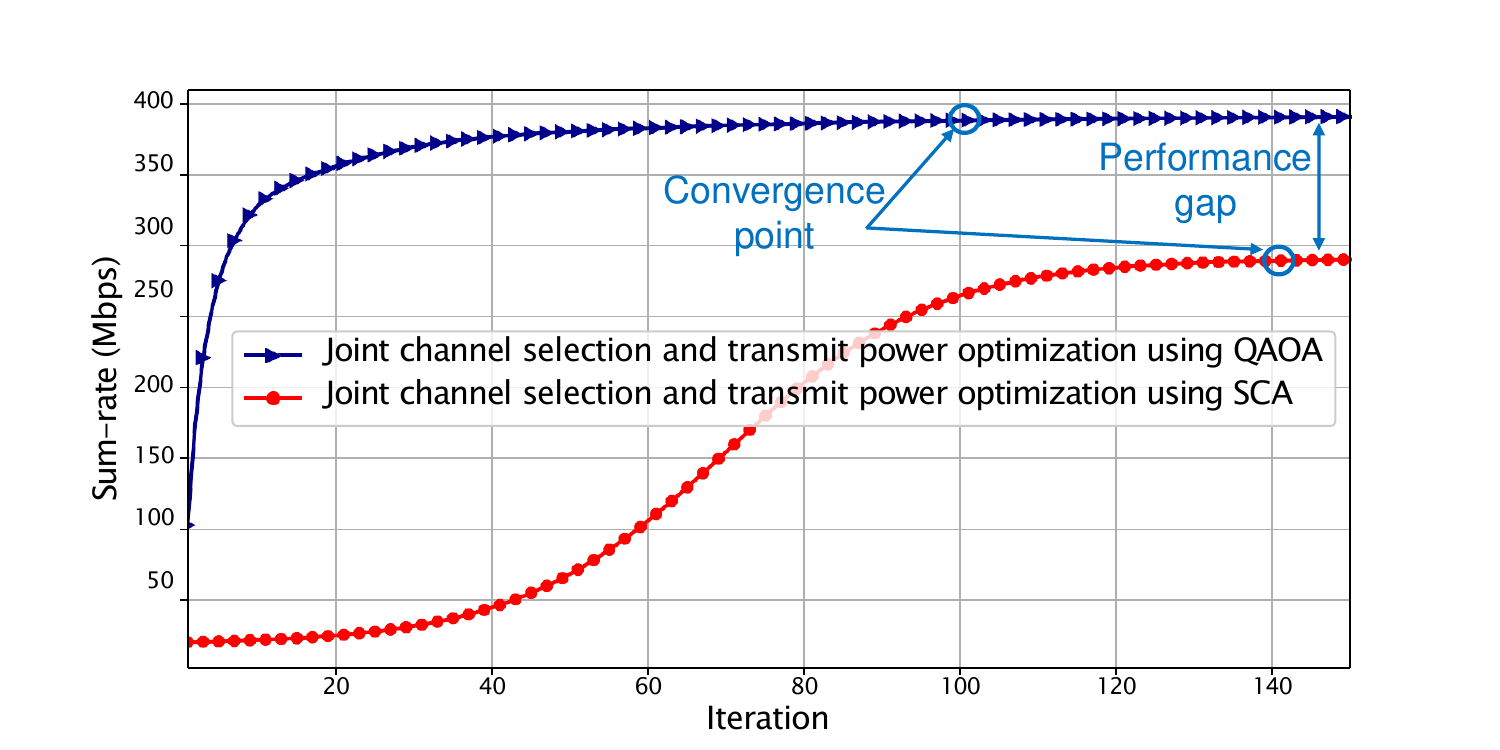}
        \caption{Comparison of joint channel selection and transmit power optimization using QAOA and SCA technique for $N = 200$ devices.}
        \label{Fig6new}
    \end{subfigure}
    \hfill
    \begin{subfigure}[b]{0.48\linewidth}
        \centering
        \includegraphics[width=\linewidth]{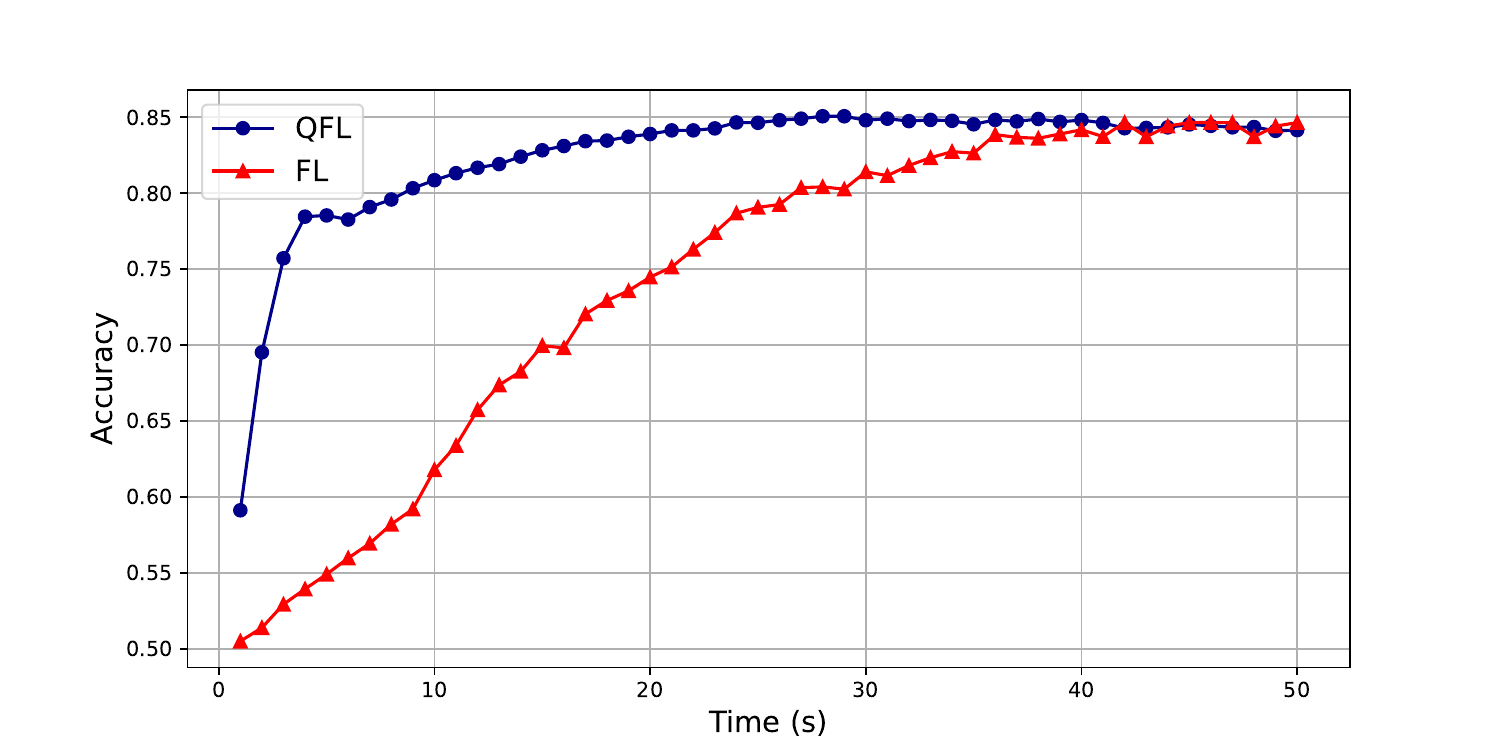}
        \caption{Comparison of training performance between QFL and FL in terms of training accuracy.}
        \label{Fig7new}
    \end{subfigure}
    \caption{\textbf{(a)}: Comparison of network optimization conducted by QAOA and SCA. \textbf{(b)}: The corresponding learning convergence of QFL and FL. The results in \textbf{(b)} where obtained on the MNIST dataset under a non-IID data setting, where each device holds data samples from only one label.}
    \label{fig:comparisonQAOA}
\end{figure*}

\subsection{System Model}
 We consider a QFL network where quantum-empowered network optimization is used for $N=200$  clients/devices who participate in collaborative model training and periodically transmit their local models to a base station (the global model is derived via a weighted average of the transmitted models, scaled by each device's number of data points).  The available bandwidth is divided into $20$ sub-channels, each with bandwidth of $180$~kHz.  In this scenario, the number of devices exceeds the number of sub-channels (a typical condition in wireless networks), which drives the need for efficient sub-channel allocation. As a result, the transmission rate of each device depends on the channel gain of its allocated sub-channel (a function of device's distance to BS considered in \textcolor{black}{[1 m, 1.8 km]} and Rician fading), its transmit power, interference caused by the other devices who use the same sub-channel, and noise.

For demonstration purposes, we aim to accelerate the convergence of the global model by facilitating faster training rounds, achieved through maximizing the sum of the data rates (sum-rate) of devices computed via conventional Shannon's formula. This is done by jointly optimizing sub-channel allocation (modeled via binary variables that capture the device-to-subchannel assignment) and transmit power (modeled via continuous variables in $[0,24]$ dBm) of devices. We address the resulting mixed-integer nonlinear programming (MINLP) problem using a QAOA approach with quadratic unconstrained binary optimization (QUBO) reformulation and block coordinate descent (BCD) \cite{choi2019tutorial}. We employ QAOA for its ability to address complex non-convex optimization problems, and our results show that QAOA outperforms the conventional successive convex approximation (SCA) approach \cite{wang2023graph}.

\subsection{Illustrative Results}

In Fig.~\ref{fig:comparisonQAOA}, we compare the performance of joint sub-channel selection and transmit power optimization tackled by QAOA and SCA techniques and the corresponding performance of QFL and FL methods. Within both QAOA and SCA, these two optimization problems were solved alternatively over 150 iterations. In Fig.~\ref{Fig6new}, we depict the sum-rate (in Mbps) vs. the number of iterations. Also, Fig.~\ref{Fig7new} displays the convergence of the corresponding QFL and FL methods.
The results demonstrate that QAOA significantly outperforms SCA in terms of both the speed of convergence and the sum-rate achieved upon convergence of the solver: QAOA achieves almost $35\%$ increase in sum-rate and converges $40\%$ faster, which results in a faster convergence of QFL contributing to the overall reduced training time in the QFL framework.


\section{Unique Challenges of QFL: To Road from Theory/Simulations to Practice}
Despite all the advantages we discussed above, we must acknowledge that realizing QFL, particularly over mobile, heterogeneous, and resource-constrained 6G wireless networks is still at the research/theory stage.  Below, we identify three interlocking dimensions where the theory-practice gap must be bridged for QFL to become a viable component of future AI-native 6G networks.

\subsection{Fragility of Quantum States in Wireless Environments}
One of the most foundational challenges in QFL is the fragility of quantum information during transmission, which is necessary for model aggregations. Unlike  ML model aggregations through classical communications, qubits, whether encoding variational parameters, quantum gradients, or entangled states, are susceptible to noise and environmental disruptions. These effects are particularly pronounced in \textit{wireless} settings, where quantum states are subject to decoherence from atmospheric scattering, Doppler shifts, and background noise, especially in mobile edge scenarios where drones, vehicles, or cellphones engage in wireless information exchange \cite{park2025entanglement}.
To address these obstacles, QFL must adopt new architectural adaptations. Promising directions include: (i) Local quantum processing combined with classical aggregation, where local quantum models (e.g., QNNs) are executed on-device, and only classical summaries (e.g., measured gradients) are communicated for aggregation;  (ii) Quantum-aware physical layer design, where new physical-layer designs enable direct inter-device quantum communication for QFL-specific tasks such as quantum state alignment (i.e., ensuring that distributed quantum devices maintain consistent reference frames, such as phase, polarization, or basis orientation); (iii) Loss-tolerant quantum communication protocols for mobile quantum links, leveraging techniques such as entanglement distillation or quantum repeaters adapted to dynamic, mobile environments where rapid link degradation is common.  These areas remain largely open for investigation.

\subsection{Incompatibility with Classical Wireless Protocols}
Today’s wireless protocol stacks are designed to transmit classical, replicable data packets. However, QFL introduces a new set of requirements: collision-free quantum state exchanges, entanglement distribution, and coherence-sensitive operations, which break the assumptions of traditional wireless protocol stacks \cite{park2025entanglement}.
To enable practical QFL deployments, we envision the evolution toward a quantum-aware communication stack, which calls for research on various subjects, including:
(i) Development of quantum-aware medium access control (Q-MAC) \cite{illiano2023quantum}, where future MAC protocols ensure reliable, low-latency access to quantum channels while preserving the fragile properties of quantum information during transmission;
(ii) Entanglement-aware information routing, where information/data routing decisions are designed to dynamically incorporate quantum-specific metrics such as entanglement fidelity and lifetime, rather than relying solely on classical notions of throughput or delay;
(iii) Cross-layer quantum-classical protocol co-design, where
codesign strategies enable adaptation of classical MAC and routing protocols  to quantum properties, such as coherence time (i.e., how long a qubit can reliably retain its quantum state) or entanglement longevity (i.e., the time window over which an entangled link remains usable for communication or computation).

\subsection{Limitations of Current Quantum Hardware for Edge Deployment}

While quantum computing hardware is advancing rapidly, today’s quantum processors, often reliant on cryogenic cooling and lab-grade isolation, remain far from deployable in edge scenarios. \textcolor{black}{In essence, the \textit{current noisy intermediate-scale quantum (NISQ)} era, characterized by limited qubit counts and short coherence times, defines the practical constraints for any near-term QFL implementation \cite{pokharel2025quantum}. To address this challenge, the feasibility of QFL relies on models/methods designed specifically for such hardware, primarily VQCs. VQCs are hybrid quantum-classical algorithms that use shallow-depth circuits, making them inherently more resilient to noise and gate errors. They can  mitigate the limitations of NISQ hardware by offloading most of the learning process to classical optimizers while using the quantum processor only where it provides a distinct advantage.}
\textcolor{black}{However, this introduces a critical trade-off in time-sensitive 6G applications: the limited availability of qubit resources and the latency involved in accessing quantum hardware may hinder the timely execution of tasks with strict deadlines, such as URLLC. As a result, in the near-term, QFL may be better suited for computationally intensive optimization problems (e.g., network planning), where achieving higher-quality solutions justifies the added processing delay.}

\textcolor{black}{The constraints of NISQ-era hardware and the current emphasis on VQC-based approaches} motivate a three-layered and incremental research roadmap: (i) \textbf{Quantum-inspired edge devices}, where research can focus on adapting classical models like tensor networks that emulate quantum phenomena as drop-in replacements in QFL pipelines; (ii) \textbf{Modular quantum coprocessors}, where hybrid architectures combine classical processors with compact quantum coprocessors for specialized subroutines; and (iii) \textbf{Energy-constrained quantum scheduling}, where device scheduling algorithms are designed for quantum-capable edge devices while accounting for their strict energy and coherence limitations. 

 \textcolor{black}{Additionally, looking ahead to the fault-tolerant quantum era, QFL systems are expected to improve significantly. Fault-tolerant quantum computers, equipped with robust quantum error correction and significantly more stable logical qubits, will unlock the potential for deep QNNs, quantum-enhanced reinforcement learning, and end-to-end quantum-classical training pipelines. These advances will enable QFL to operate over more expressive quantum models with superior generalization capabilities and allow low-latency, on-device quantum inference even in noisy wireless environments. In preparation for this paradigm shift, it will be critical to (i) \textbf{Develop scalable quantum data encoding schemes}, such as amplitude encoding and quantum analog-to-digital/digital-to-analog converters (QADC/QDAC), that can support high-dimensional feature inputs while respecting practical limits on gate depth and error thresholds; (ii) \textbf{Design efficient, fault-tolerant circuit synthesis pipelines for core QFL components}, including quantum kernel methods, QNN layers, and variational ansätze, tailored to the logical qubit layout of target architectures; and (iii) \textbf{Enable cross-layer co-design frameworks} that jointly optimize quantum circuit execution, wireless communication scheduling, and federated aggregation strategies in the presence of dynamic network conditions and quantum noise.}

\section{Conclusion}
In this paper, we investigated how integration of QFL in AI-native
6G networks can lead to overcoming key challenges, including
edge intelligence, network optimization, and security.
Moreover, we conducted a case study where QAOA is used
to optimize channel selection and transmit power in 6G QFL networks. Simulation results revealed that QAOA outperforms
the classical SCA technique by delivering higher sum-rates
and achieving faster model convergence, demonstrating the
potential advantages of QFL in AI-native 6G networks. Finally,
we discussed practical challenges for QFL deployment,
including quantum state fragility, protocol incompatibility, and
hardware constraints, suggesting future research pathways to
advance QFL in 6G wireless networks.

\bibliographystyle{ieeetr}
\bibliography{references}

\begin{IEEEbiographynophoto}{ Shaba Shaon} is currently pursuing her Ph.D. at the College of Engineering, The University of Alabama in Huntsville, Huntsville, AL, USA. She is with the Networking, Intelligence, and Security Research Lab at the same institution. Her research has been submitted to renowned IEEE journals and conferences. Her research interests focus on quantum federated learning, wireless communication, networking, and optimization algorithms.
\end{IEEEbiographynophoto}

\vspace{-33pt}
\begin{IEEEbiographynophoto}{ Md Raihan Uddin} is a master’s student in computer engineering at the College of Engineering, The University of Alabama in Huntsville, USA, where he also serves as a researcher at the Networking, Intelligence, and Security Lab. He is the Graduate Student Member of IEEE. He has developed a robust foundation in multiple aspects of technology, particularly in machine learning, artificial intelligence, quantum learning, privacy and security, and wireless communication.
\end{IEEEbiographynophoto}

\vspace{-30pt}
\begin{IEEEbiographynophoto}{ Dinh C. Nguyen} is an assistant professor at the Department of Electrical and Computer Engineering, The University of Alabama in Huntsville, USA. He worked as a postdoctoral research associate at Purdue University, USA from 2022 to 2023. He obtained the Ph.D. degree in computer science from Deakin University, Australia in 2021. His current research interests include federated machine learning, Internet of Things, wireless networking, and security.   He received the Best Editor Award from IEEE Open Journal of Communications Society in 2023.
\end{IEEEbiographynophoto}

\vspace{-30pt}
\begin{IEEEbiographynophoto}{Seyyedali Hosseinalipour} received the B.S. degree in electrical engineering from Amirkabir University of Technology, Tehran, Iran, in 2015 with high honor and top-rank recognition. He then received the M.S. and Ph.D. degrees in electrical engineering from North Carolina State University, NC, USA, in 2017 and 2020, respectively; and was a postdoctoral researcher at Purdue University, IN, USA from 2020 to 2022. He was the recipient of the ECE Doctoral Scholar of the Year Award (2020) and ECE Distinguished Dissertation Award (2021) at NC State University; and Students’ Choice Teaching Excellence Award (2023) at University at Buffalo–SUNY. Furthermore, he was the first author of a paper published in IEEE/ACM Transactions on Networking that received the 2024 IEEE Communications Society William Bennett Prize. He has served as the TPC Co-Chair of workshops/symposiums related to machine learning and edge computing for IEEE INFOCOM, GLOBECOM, ICC, CVPR, ICDCS, SPAWC, WiOpt, and VTC. He has also served as the guest editor of IEEE Internet of Things Magazine for the special issue on Federated Learning for Industrial Internet of Things (2023). Since Feb. 2025, he has been serving as an Associate Editor for the IEEE Transactions on Signal and Information Processing over Networks. His research interests include the analysis of modern wireless networks, synergies between machine learning methods and fog/edge computing systems, distributed/federated machine learning, and network optimization.
\end{IEEEbiographynophoto}

\vspace{-30pt}
\begin{IEEEbiographynophoto}{Dusit Niyato} is a professor in the College of Computing and Data Science at Nanyang Technological University, Singapore. He received B.Eng. from King Mongkuts Institute of Technology Ladkrabang (KMITL), Thailand in 1999 and Ph.D. in Electrical and Computer Engineering from the University of Manitoba, Canada in 2008. His research interests are in the areas of sustainability, edge intelligence, decentralized machine learning, and incentive mechanism design.
\end{IEEEbiographynophoto}

\vspace{-30pt}
\begin{IEEEbiographynophoto}{Octavia A. Dobre} is a Professor and Tier-1 Canada Research Chair with Memorial University, Canada. Her research interests encompass wireless communication and networking technologies, as well as optical and underwater communications. She has (co-)authored over 600 publications in these areas, and obtained ten Best Paper Awards including the IEEE Communications Society Heinrich Hertz Award. Dr. Dobre serves as the VP Publications of the IEEE Communications Society. She was the inaugural Editor-in-Chief (EiC) of the IEEE Open Journal of the Communications Society and the EiC of the IEEE Communications Letters. Dr. Dobre is an elected member of the European Academy of Sciences and Arts, a Fellow of the Engineering Institute of Canada, a Fellow of the Canadian Academy of Engineering, and a Fellow of the Royal Society of Canada.
\end{IEEEbiographynophoto}

\vfill

\end{document}